\documentclass[final,1p]{elsarticle}
\usepackage[T1]{fontenc}
\usepackage{graphics}
\usepackage{verbatim}
\usepackage{amssymb}
\usepackage{amsmath}
\usepackage{graphicx}
\usepackage{bm}
\usepackage{xcolor}
\usepackage[colorlinks,breaklinks]{hyperref}
\usepackage{tikz}
\makeatletter
\usepackage[normalem]{ulem}

\providecommand{\LyX}{L\kern-.1667em\lower.25em\hbox{Y}\kern-.125emX\@}

\journal{Computer Physics Communications}

\makeatother
\begin{document}

\newcommand{\etal}{\textit{et al.~}}

\newcommand{\rev}[1]{#1} 
\newcommand{\revb}[1]{#1} 

\newcommand{\be}{\begin{eqnarray}}
\newcommand{\ee}{\end{eqnarray}}
\newcommand{\bi}{\begin{itemize}}
\newcommand{\ei}{\end{itemize}}
\newcommand{\bwt}{\begin{widetext}}
\newcommand{\ewt}{\end{widetext}}

\newcommand{\vecr}{{\vec r}}
\newcommand{\vecR}{{\vec R}}
\newcommand{\pn}{p\text{-}n}
\newcommand{\half}{\frac{1}{2}}
\newcommand{\dif}{{\mathrm{d}}}

\newcommand{\bk}{{\mathbf k}}
\newcommand{\bK}{{\mathbf K}}
\newcommand{\bR}{{\mathbf R}}
\newcommand{\br}{{\mathbf r}}

\newcommand{\threej}[6]{\begin{pmatrix}#1&#2&#3\\#4&#5&#6\end{pmatrix}}
\newcommand{\threejz}[3]{\begin{pmatrix}#1&#2&#3\\0&0&0\end{pmatrix}}
\newcommand{\sixj}[6]{\begin{Bmatrix}#1&#2&#3\\#4&#5&#6\end{Bmatrix}}
\newcommand{\ninej}[9]{\begin{Bmatrix}#1&#2&#3\\#4&#5&#6\\#7&#8&#9\end{Bmatrix}}
\newcommand{\reduced}[3]{\langle#1 \| #2 \| #3\rangle}
\newcommand{\coords}{\vec{R},\vec{r},\xi}
\newcommand{\phic}{$\phi_{\bk_I; I \mu; s \sigma}^{(+)}(\br,\xi_v,\xi_c)$}

\begin{frontmatter}
\title{HPRMAT: A high-performance R-matrix solver with GPU acceleration for coupled-channel problems in nuclear physics\footnote{Accepted manuscript. Published as J. Lei, Comput. Phys. Commun. \textbf{329}, 110379 (2026), \url{https://doi.org/10.1016/j.cpc.2026.110379}. \copyright{} 2026. This manuscript version is made available under the CC BY-NC-ND 4.0 license, \url{https://creativecommons.org/licenses/by-nc-nd/4.0/}.}}
\date{}  
\author[Tongji]{Jin Lei}
\ead{jinl@tongji.edu.cn} 
\address[Tongji]{School of Physics Science and Engineering, Tongji University, Shanghai 200092, China}




\begin{abstract}
I present HPRMAT, \rev{a self-contained, high-performance R-matrix solver framework for coupled-channel scattering calculations in nuclear physics. It provides the full R-matrix propagation machinery with the same user-supplied-potential interface as standard R-matrix packages, and is additionally a drop-in replacement for the linear algebra routines of Descouvemont's package. It employs direct linear equation solving with optimized libraries instead of traditional matrix inversion, achieving significant performance improvements.} The package provides four solver backends: (1) double-precision LU factorization, (2) mixed-precision arithmetic with iterative refinement, (3) a Woodbury formula approach exploiting the kinetic-coupling matrix structure, and (4) GPU acceleration. Benchmark calculations demonstrate that the GPU solver achieves \rev{about 15$\times$ speedup over the optimized CPU direct solver, and 41$\times$ over the legacy inversion-based code, at $N=25600$}. The mixed-precision strategy is particularly effective on consumer GPUs (e.g., NVIDIA RTX 3090/4090), where single-precision throughput exceeds double-precision by a factor of 64:1; by performing the factorization in single precision\rev{, with iterative refinement available to recover full double-precision accuracy where needed}, HPRMAT overcomes the poor FP64 performance of consumer hardware while \rev{retaining the accuracy required for cross-section calculations}. This makes large-scale \rev{continuum-discretized coupled-channels (CDCC)} and coupled-channel calculations accessible to researchers using standard desktop workstations, without requiring expensive data-center GPUs. \rev{A single consumer card already handles the largest problems encountered in practice, and for the occasional case that exceeds one card a tested multi-GPU back-end distributes the single-precision factorization across several GPUs and recovers full double-precision accuracy through host-side iterative refinement.} CPU-only solvers provide up to about 7$\times$ speedup for large matrices (system-dependent) through optimized libraries and algorithmic improvements. \rev{All solvers reproduce the reference results well within experimental uncertainties: the double-precision and CPU mixed-precision solvers agree to machine precision, and the GPU single-precision solver to the $10^{-3}$ level in cross sections (with optional iterative refinement available for higher accuracy), validated against Descouvemont's reference code (Comput.\ Phys.\ Commun.\ 200, 199--219 (2016)).} HPRMAT provides interfaces for Fortran, C, Python, and Julia.

{\bf PROGRAM SUMMARY} \\
{\em Program Title: HPRMAT}                                          \\
{\em CPC Library link to program files:} \url{https://doi.org/10.17632/gcnpfnz3w6.1} \\
{\em Developer's repository link:} \url{https://github.com/jinleiphys/HPRMAT} \\
{\em Licensing provisions:} MIT license  \\
{\em Programming language:} Fortran 90/95, CUDA C                                  \\
{\em Nature of problem:} Solving coupled-channel Schr\"{o}dinger equations for nuclear scattering using the R-matrix method with Lagrange-Legendre basis functions. The computational bottleneck is solving large dense complex linear systems arising from the discretization of the internal region. \\
{\em Solution method:} Direct solution of the linear system using optimized LAPACK/cuSOLVER routines, with options for mixed-precision arithmetic and Woodbury formula decomposition exploiting the kinetic-coupling matrix structure. \\
{\em Additional comments including restrictions and unusual features:} GPU solver requires NVIDIA GPU with CUDA support (tested with CUDA 11.5). The in-core single-GPU solver runs up to $N\approx25600$ on a 24~GB card; the cusolverMg-based multi-GPU back-end, which keeps the single-precision factorization on the device and the double-precision matrix on the host, reaches $N=51200$ already on one such card and has been benchmarked up to $N=76800$ on four RTX~3090 GPUs (Section ``Scalability and multi-GPU support''). The code is compatible with Descouvemont's R-matrix package interface. In addition to the native Fortran API, C, Python, and Julia bindings are provided.\\

\end{abstract}


\end{frontmatter}

\section{Introduction}
\label{sec:intro}

The R-matrix method~\cite{Lane1958} has been a cornerstone of nuclear reaction theory for over six decades, providing a powerful framework for describing resonance phenomena and scattering processes in nuclear physics. By dividing configuration space into internal and external regions at a channel radius $a$, the method provides a natural pole expansion \rev{of the $R$-matrix in terms of resonance energies and reduced widths, from which the scattering matrix is subsequently obtained by matching to the asymptotic solutions at $a$}.

In microscopic R-matrix calculations~\cite{Descouvemont2010,Baye2015}, the internal wave function is expanded in a complete set of $\mathcal{L}^2$ (square-integrable) basis functions---typically Lagrange-Legendre functions derived from Gauss-Legendre quadrature. This makes the R-matrix method a \rev{\emph{bound-state-like approach to continuum problems}}~\cite{Johnson2020,Carbonell2014}: \rev{the scattering problem is formulated in a finite internal region, and the addition of the Bloch operator at the channel boundary ensures the Hermiticity of the internal-region Hamiltonian while retaining the coupling to the external continuum.} The continuum nature of the scattering problem is recovered by matching to the known asymptotic solutions at the channel radius. This approach combines the computational advantages of bound-state techniques (finite matrix diagonalization, variational principles) with the ability to describe scattering and resonance phenomena.

While direct integration methods such as the Numerov algorithm~\cite{Thorlacius1987} are computationally faster for simple single-channel scattering problems, the microscopic R-matrix approach offers distinct advantages for coupled-channel calculations. For multi-channel problems, the R-matrix method solves a single linear system for all channels simultaneously, whereas Numerov requires integrating each channel separately with careful treatment of channel coupling. The method also handles threshold cusps and near-threshold behavior naturally through the analytic properties of the R-matrix, where Numerov integration may require extremely fine step sizes. Furthermore, the $\mathcal{L}^2$ basis expansion avoids the numerical instabilities that can arise in Numerov integration for deeply bound or highly oscillatory wave functions in the internal region. These features make the microscopic R-matrix approach particularly well-suited for problems involving strong channel coupling near thresholds. \rev{I note that this comparison is specifically with \emph{direct} Numerov integration. Other multichannel propagators widely used in atomic and molecular scattering, such as the log-derivative propagator~\cite{Johnson1973} and the Light-Walker R-matrix propagator~\cite{LightWalker1976}, treat all coupled channels simultaneously through diagonalizations or the solution of linear systems, and mitigate some of the stability issues of direct Numerov integration near thresholds.}

Modern applications of the coupled-channel method increasingly require the treatment of many channels. Examples include heavy-ion fusion reactions with rotational and vibrational couplings~\cite{Hagino1999,Hagino2022}, and elastic scattering and breakup reactions of weakly-bound nuclei described by the continuum-discretized coupled-channels (CDCC) method~\cite{Austern1987}. In such calculations, the number of coupled channels can easily exceed 50, leading to linear systems with matrix dimensions $N = n_{\rm ch} \times n_{\rm lag}$ reaching several thousand, where $n_{\rm ch}$ is the number of channels and $n_{\rm lag}$ is the number of Lagrange basis functions per channel.

The CDCC method represents a particularly demanding application. To describe the breakup continuum of a weakly-bound projectile, the continuum must be discretized into ``bin'' states up to a cutoff energy, with each bin contributing multiple partial waves. Halo nuclei such as $^{11}$Be (a single-neutron halo with $^{10}$Be core) and $^{6}$He (a two-neutron halo) are especially challenging due to their extended spatial distributions and low breakup thresholds. Extended CDCC (XCDCC) calculations that include core excitations~\cite{Summers2006} are particularly demanding. A typical XCDCC calculation for $^{11}$Be scattering~\cite{Pesudo2017} requires: (1) bound states and continuum bins for each projectile spin-parity $J^\pi$ from $1/2^\pm$ up to $15/2^\pm$; (2) multiple energy bins per $J^\pi$ (typically 10--15 bins at lower energies, 4--5 at higher energies); (3) partial waves up to $l_{\rm max} \sim 9$ for the core-valence relative motion. This can generate $n_{\rm ch} \sim 200$--$400$ projectile pseudostates. With $n_{\rm lag} = 40$--$80$ Lagrange basis functions needed for convergence, matrix dimensions of $N = 10{,}000$--$30{,}000$ are routinely required. This motivates the focus on large-matrix performance in the present work.

The computational bottleneck in R-matrix calculations is the solution of the resulting complex linear system. Traditional implementations, such as the widely-used code by Descouvemont~\cite{Descouvemont2016}, employ matrix inversion to obtain the R-matrix from the Hamiltonian matrix. While mathematically equivalent to solving a linear system, matrix inversion has several disadvantages: it requires computing and storing the full inverse matrix, is numerically less stable than direct linear solvers, and cannot take advantage of modern numerical algorithms optimized for linear systems.

Several R-matrix codes exist for different applications. \rev{In atomic and molecular physics, the UK PRMAT code~\cite{Noble2002} employs MPI parallelization and ScaLAPACK for distributed computing, and the UKRmol+ suite~\cite{Masin2020} is a widely used implementation for electron, positron, and photon collisions with molecules; GPU acceleration has also begun to appear in that domain, for example in a recent R-matrix treatment of H$_2$O\,+\,H inelastic collisions~\cite{GarciaVazquez2026}. In nuclear physics, however, to my knowledge no publicly available R-matrix code exploits GPU acceleration or modern mixed-precision techniques.} \revb{The present work fills this gap.}

In this paper, I present HPRMAT (High-Performance R-MATrix), \rev{a self-contained, high-performance R-matrix solver framework for coupled-channel scattering calculations. HPRMAT provides the complete propagation machinery (Lagrange-Legendre mesh, kinetic and Bloch operators, collision-matrix construction with Coulomb asymptotics, and wave-function reconstruction), together with four interchangeable solver backends and interfaces for Fortran, C, Python, and Julia. It adopts the same user-supplied-potential interface as Descouvemont's standard reference package~\cite{Descouvemont2016}: the physics setup (potential construction, channel coupling, boundary conditions, Buttle correction if needed, etc.) is supplied by the user, exactly as in that package. As an additional deployment convenience, HPRMAT is drop-in compatible with Descouvemont's package, so its existing users can adopt the new solvers without modifying their application codes.} \revb{HPRMAT represents the current \emph{state-of-the-art} in R-matrix solver performance among publicly available codes for nuclear physics applications. In my benchmarks using a consumer-grade NVIDIA RTX 3090 GPU, the GPU solver achieves about 15$\times$ speedup over the optimized CPU direct solver (and 41$\times$ over the legacy inversion-based code) at $N = 25600$; this advantage would be larger still on more powerful consumer GPUs with greater FP32 throughput.} HPRMAT addresses the computational bottleneck through several innovations:
\begin{enumerate}
    \item \textbf{Adoption of direct linear solvers.} While solving $\mathbf{A}\mathbf{x}=\mathbf{b}$ directly rather than computing $\mathbf{x}=\mathbf{A}^{-1}\mathbf{b}$ is standard practice in numerical linear algebra, many legacy nuclear physics codes---including Descouvemont's widely-used package---still employ explicit matrix inversion. HPRMAT implements direct LU factorization, bringing this well-established best practice to the R-matrix community.
    \item \textbf{Optimized BLAS library integration.} I utilize OpenBLAS~\cite{OpenBLAS}, a highly optimized open-source BLAS implementation that exploits multi-threading and SIMD vectorization, providing substantial speedups over reference LAPACK on modern multi-core CPUs.
    \item \textbf{GPU acceleration via NVIDIA cuSOLVER.} I provide a GPU-accelerated solver that achieves up to about 20$\times$ speedup over optimized CPU solvers for large matrices on consumer-grade GPUs.
    \item \textbf{Mixed-precision arithmetic.} I exploit the favorable FP32:FP64 performance ratio (64:1 on RTX 3090) by performing LU factorization in single precision while maintaining double-precision accuracy through iterative refinement.
    \item \textbf{Woodbury formula optimization.} I develop a solver that exploits the specific block structure of the coupled-channel Hamiltonian matrix, where diagonal blocks are full matrices (kinetic energy) and off-diagonal blocks are diagonal (coupling potentials).
    \item \textbf{Multi-language interfaces.} I provide bindings for C, Python, and Julia, enabling seamless integration into modern scientific workflows beyond Fortran.
\end{enumerate}

A key design principle of HPRMAT is \emph{drop-in compatibility} with Descouvemont's R-matrix package~\cite{Descouvemont2016}. The subroutine interfaces, calling conventions, and data structures are kept as close as possible to the original implementation. Users of Descouvemont's code can switch to HPRMAT by simply replacing the relevant source files and relinking, without modifying their existing application codes. This design choice ensures that the performance improvements are immediately accessible to the existing user community. All solvers have been validated against Descouvemont's reference implementation using all five standard test cases from his package: $\alpha$+$^{208}$Pb optical model scattering, nucleon-nucleon scattering with the Reid soft-core potential, $^{16}$O+$^{44}$Ca coupled-channel scattering, $^{12}$C+$\alpha$ inelastic scattering, and the non-local Yamaguchi potential.

This paper is organized as follows. Section~\ref{sec:theory} briefly reviews the R-matrix formalism and the structure of the resulting linear system. Section~\ref{sec:methods} describes the four solver algorithms implemented in HPRMAT. Section~\ref{sec:implementation} discusses implementation details including the Fortran-CUDA interface and BLAS library integration. Section~\ref{sec:results} presents benchmark results on three different hardware platforms. Section~\ref{sec:conclusion} summarizes the findings and discusses future directions.

\section{Theoretical Background}
\label{sec:theory}

The R-matrix method~\cite{Descouvemont2010,Baye2015} provides a framework for solving the Schr\"{o}dinger equation for scattering problems by dividing configuration space at a channel radius $r = a$. In the internal region ($r < a$), the wave function is expanded in a complete basis set. In the external region ($r > a$), the wave function is expressed in terms of known asymptotic solutions that are matched to the internal solution at the boundary.

For a system with $n_{\rm ch}$ coupled channels, the radial Schr\"{o}dinger equation takes the form
\be
\left[ -\frac{\hbar^2}{2\mu} \frac{d^2}{dr^2} + \frac{\ell_\alpha(\ell_\alpha+1)\hbar^2}{2\mu r^2} + V_{\alpha\alpha}(r) - E \right] \psi_\alpha(r) + \sum_{\alpha' \neq \alpha} V_{\alpha\alpha'}(r) \psi_{\alpha'}(r) = 0,
\label{eq:cc_schrodinger}
\ee
where $\alpha$ labels the channels (including internal quantum numbers and orbital angular momentum $\ell_\alpha$), $\mu$ is the reduced mass, and $V_{\alpha\alpha'}(r)$ are the diagonal ($\alpha = \alpha'$) and coupling ($\alpha \neq \alpha'$) potentials.

Following Baye~\cite{Baye2015} and Descouvemont~\cite{Descouvemont2016}, the internal wave function is expanded in Lagrange-Legendre basis functions $\{f_i(r)\}_{i=1}^{n_{\rm lag}}$ defined on a mesh of Gauss-Legendre quadrature points $\{r_i\}$ with corresponding weights $\{\lambda_i\}$ on the interval $(0, a)$. These basis functions satisfy the Lagrange property $f_i(r_j) = \delta_{ij}/\sqrt{\lambda_j}$, which allows potential matrix elements to be evaluated as simple function values at mesh points rather than numerical integrals. The wave function in channel $\alpha$ is expanded as $\psi_\alpha(r) = \sum_{i=1}^{n_{\rm lag}} c_{\alpha i} f_i(r)$, leading to a linear system for the expansion coefficients $c_{\alpha i}$.

Substituting the basis expansion into Eq.~(\ref{eq:cc_schrodinger}) yields a linear system
\be
\mathbf{M} \cdot \mathbf{c} = \mathbf{b},
\label{eq:linear_system}
\ee
where $\mathbf{M}$ is a complex matrix of dimension $N \times N$ with $N = n_{\rm ch} \times n_{\rm lag}$, $\mathbf{c}$ is the coefficient vector, and $\mathbf{b}$ \rev{is the surface (Bloch) source term. Rather than the asymptotic boundary conditions themselves, its entries are the boundary values of the Lagrange basis functions at the channel radius $a$ [the $q_2(r_i)$ of Sec.~\ref{sec:methods}], which arise from the Bloch operator and encode the matching of the internal solution to the external region}.

The matrix $\mathbf{M}$ has a characteristic block structure:
\be
\mathbf{M} = \begin{pmatrix}
\mathbf{K}_1 + \mathbf{D}_{11} & \mathbf{D}_{12} & \cdots & \mathbf{D}_{1,n_{\rm ch}} \\
\mathbf{D}_{21} & \mathbf{K}_2 + \mathbf{D}_{22} & \cdots & \mathbf{D}_{2,n_{\rm ch}} \\
\vdots & \vdots & \ddots & \vdots \\
\mathbf{D}_{n_{\rm ch},1} & \mathbf{D}_{n_{\rm ch},2} & \cdots & \mathbf{K}_{n_{\rm ch}} + \mathbf{D}_{n_{\rm ch},n_{\rm ch}}
\end{pmatrix},
\label{eq:block_structure}
\ee
where:
\begin{itemize}
    \item $\mathbf{K}_\alpha$ is the kinetic energy matrix for channel $\alpha$, a \emph{full} $n_{\rm lag} \times n_{\rm lag}$ matrix including the Bloch operator boundary term;
    \item $\mathbf{D}_{\alpha\alpha'}$ is the coupling potential matrix. For \emph{local potentials}, this matrix is \emph{diagonal} in the Lagrange basis due to the Lagrange property:
    \be
    (D_{\alpha\alpha'})_{ij} = V_{\alpha\alpha'}(r_i) \delta_{ij}.
    \ee
    For non-local potentials, $\mathbf{D}_{\alpha\alpha'}$ becomes a full matrix, and the sparsity structure discussed below no longer applies.
\end{itemize}

\begin{figure}[htbp]
\centering
\includegraphics[width=0.7\textwidth]{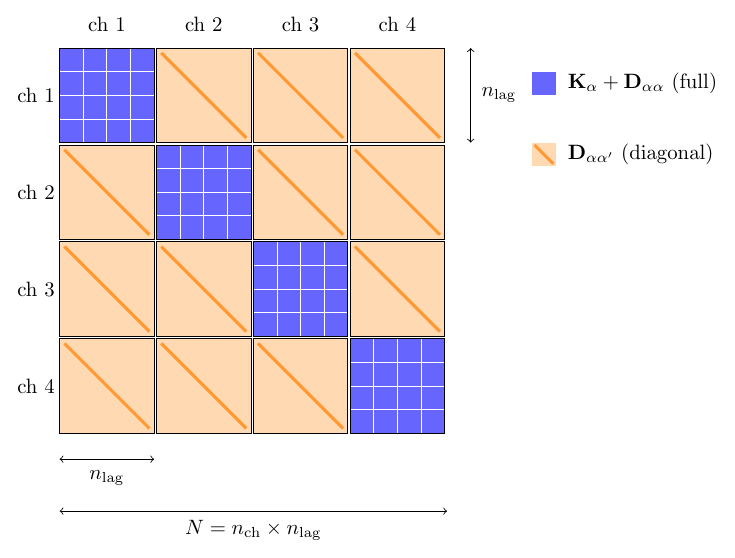}
\caption{Schematic structure of the R-matrix Hamiltonian for a 4-channel problem ($n_{\rm ch} = 4$) with \emph{local potentials}. Diagonal blocks (blue, dense pattern) contain the full kinetic energy plus Bloch operator matrices $\mathbf{K}_\alpha$ and diagonal potentials; off-diagonal blocks (orange, diagonal line) are diagonal coupling matrices $\mathbf{D}_{\alpha\alpha'}$. Each block has dimension $n_{\rm lag} \times n_{\rm lag}$, giving total matrix dimension $N = n_{\rm ch} \times n_{\rm lag}$. For non-local potentials, all blocks become full matrices.}
\label{fig:matrix_structure}
\end{figure}

For local potentials, this structure is illustrated schematically in Fig.~\ref{fig:matrix_structure}. Although the matrix appears sparse (only $\sim$3\% of elements are non-zero for typical problems with $n_{\rm ch} \sim 50$ and $n_{\rm lag} \sim 100$), this sparsity is \emph{deceptive} and cannot be exploited by standard sparse or iterative solvers. The fundamental difficulty is that the matrix has strong coupling in \emph{two directions simultaneously}: (1) full dense coupling within each channel through the kinetic energy matrix $\mathbf{K}_\alpha$, and (2) full coupling between all channels at each radial point through the potential matrices $\mathbf{D}_{\alpha\alpha'}$. During LU factorization, the off-diagonal blocks rapidly fill in, destroying the initial sparsity pattern.

I tested numerous alternative approaches during development:
\begin{itemize}
    \item \textbf{Block Gaussian elimination}~\cite{Anderson1999}: Correctly handles the fill-in during elimination, but the computational cost is identical to dense LU since all blocks become full matrices.
    \item \textbf{Block Thomas algorithm}~\cite{Golub2013}: Attempted to exploit the block structure by treating the matrix as block-tridiagonal, but the approximation error exceeded 10\%.
    \item \textbf{Block Jacobi preconditioned GMRES}~\cite{Saad1986,Saad2003}: Failed to converge due to strong inter-channel coupling; the block-diagonal preconditioner is too weak, with residuals stagnating at 3--12\%.
    \item \textbf{ILU(0)-preconditioned GMRES}~\cite{Saad2003}: Converged but required over 600 iterations, making it 50$\times$ slower than direct methods.
    \item \textbf{UMFPACK sparse direct solver}~\cite{Davis2004}: Provided only 5--10\% speedup for small channel numbers ($n_{\rm ch} < 35$) and became slower than dense methods for larger problems due to fill-in.
    \item \textbf{Woodbury formula with radial-diagonal base}~\cite{Hager1989}: Decomposed $\mathbf{M} = \mathbf{C} + \mathbf{I}_{n_{\rm ch}} \otimes \mathbf{K}$ where $\mathbf{C}$ contains the coupling potentials as $n_{\rm lag}$ blocks of size $n_{\rm ch} \times n_{\rm ch}$. Although $\mathbf{C}^{-1}$ is easy to compute (invert $n_{\rm lag}$ small matrices), the Schur complement after the Woodbury transformation remains $N \times N$ and dense, providing no computational advantage.
\end{itemize}
These negative results motivated the focus on optimizing dense direct solvers described in Section~\ref{sec:methods}. The key insight is that for this problem class, the best strategy is not to fight the fill-in but to embrace dense methods and accelerate them through optimized libraries, mixed-precision arithmetic, and GPU offloading.

\section{Numerical Methods}
\label{sec:methods}

HPRMAT implements four solver backends, allowing users to choose the optimal algorithm for their hardware and accuracy requirements. All solvers take the matrix $\mathbf{M}$ and right-hand side vector $\mathbf{b}$ as input and return the solution vector $\mathbf{c}$.

A critical factor for CPU performance is the choice of BLAS library. All CPU solvers in HPRMAT are designed to use OpenBLAS~\cite{OpenBLAS}, an optimized open-source implementation that provides multi-threaded BLAS/LAPACK routines. Unlike the reference LAPACK library, OpenBLAS exploits modern CPU features including SIMD vectorization (AVX/AVX-512) and multi-core parallelism, achieving near-peak floating-point performance. This is essential for the observed speedups, as the same algorithm linked against reference LAPACK would be significantly slower.

\subsection{Dense Solver (Type 1)}

The reference solver uses the ZGESV routine from OpenBLAS, which provides a highly optimized implementation of the standard LAPACK interface. ZGESV performs LU factorization with partial pivoting followed by forward and backward substitution:
\be
\mathbf{M} = \mathbf{P} \cdot \mathbf{L} \cdot \mathbf{U},
\ee
where $\mathbf{P}$ is a permutation matrix, $\mathbf{L}$ is lower triangular with unit diagonal, and $\mathbf{U}$ is upper triangular.

The R-matrix calculation requires solving for multiple right-hand sides simultaneously. For each channel $\alpha$, the right-hand side vector $\mathbf{b}_\alpha$ contains the boundary basis function values $q_2(r_i)$ at the Lagrange mesh points. The algorithm proceeds as follows:
\begin{enumerate}
    \item Copy the Hamiltonian matrix $\mathbf{M}$ to a work array (ZGESV overwrites the input);
    \item Construct the right-hand side matrix $\mathbf{B}$ of dimension $N \times n_{\rm ch}$, where each column $\alpha$ contains $q_2(r_i)$ in the block corresponding to channel $\alpha$;
    \item Call ZGESV to solve $\mathbf{M} \mathbf{X} = \mathbf{B}$ for all $n_{\rm ch}$ right-hand sides simultaneously;
    \item Extract R-matrix elements: $R_{\alpha\alpha'} = \sum_{i=1}^{n_{\rm lag}} q_2(r_i) \cdot X_{i+(\alpha-1)n_{\rm lag}, \alpha'}$.
\end{enumerate}

The computational complexity is $O(N^3)$ for the LU factorization and $O(N^2 \cdot n_{\rm ch})$ for the solve phase. This solver achieves machine precision ($\sim 10^{-18}$ for double-precision complex arithmetic) and serves as the accuracy reference for other solvers.

Compared to the matrix inversion approach used in Descouvemont's code~\cite{Descouvemont2016}, the direct solve has several advantages:
\begin{itemize}
    \item Lower memory requirements (no need to store the full inverse matrix);
    \item Better numerical stability (avoids amplification of rounding errors in explicit inversion);
    \item Smaller computational constant factor (despite both being $O(N^3)$).
\end{itemize}

\subsection{Mixed-Precision Solver (Type 2)}

Modern processors execute single-precision (FP32) operations faster than double-precision (FP64). On CPUs, the speedup is typically 2$\times$ due to wider SIMD registers. On consumer GPUs, the ratio can be much larger---the NVIDIA RTX 3090 achieves 35 TFLOPS in FP32 but only 0.56 TFLOPS in FP64, a ratio of 64:1.

The mixed-precision solver exploits this asymmetry by performing the expensive $O(N^3)$ LU factorization in single precision (CGETRF) while maintaining double-precision accuracy through iterative refinement. The initial single-precision solution is refined by computing the residual $\mathbf{r} = \mathbf{b} - \mathbf{M}\mathbf{x}$ in double precision and solving for a correction using the stored single-precision LU factors. With 1--2 refinement iterations, the final accuracy reaches double precision (\rev{$\sim 10^{-15}$}), while the computational cost is dominated by the fast single-precision factorization.

\subsection{Woodbury-Kinetic Solver (Type 3)}

\textbf{Note:} This solver is specifically designed for \emph{local potentials} and exploits the diagonal structure of the off-diagonal coupling blocks. For non-local potentials, where all blocks are full matrices, use Type 1, 2, or 4 solvers instead.

The Woodbury matrix identity~\cite{Woodbury1950} provides an efficient method to solve linear systems when the matrix can be expressed as a structured perturbation of an easily invertible base matrix:
\be
(\mathbf{A} + \mathbf{U}\mathbf{C}\mathbf{V})^{-1} = \mathbf{A}^{-1} - \mathbf{A}^{-1}\mathbf{U}(\mathbf{C}^{-1} + \mathbf{V}\mathbf{A}^{-1}\mathbf{U})^{-1}\mathbf{V}\mathbf{A}^{-1}. \label{eq:woodbury}
\ee

The R-matrix Hamiltonian has a specific block structure that can be exploited. Recalling Eq.~(\ref{eq:block_structure}), the diagonal blocks $\mathbf{K}_\alpha + \mathbf{D}_{\alpha\alpha}$ contain the kinetic energy (full matrices) plus diagonal potentials, while off-diagonal blocks $\mathbf{D}_{\alpha\alpha'}$ ($\alpha \neq \alpha'$) are purely diagonal matrices representing channel coupling. I decompose the matrix as:
\be
\mathbf{M} = \mathbf{K}_{\rm diag} + \mathbf{V}_{\rm coupling},
\ee
where $\mathbf{K}_{\rm diag}$ is block-diagonal with blocks $(\mathbf{K}_\alpha + \mathbf{D}_{\alpha\alpha})$, and $\mathbf{V}_{\rm coupling}$ contains only the off-diagonal coupling terms.

\rev{To make the connection to Eq.~(\ref{eq:woodbury}) explicit: the base matrix plays the role of $\mathbf{A}$ and is $\mathbf{K}_{\rm diag}$, which is cheap to handle because it is block-diagonal and the kinetic block $\mathbf{K}$ is real and shared across all channels. The coupling $\mathbf{V}_{\rm coupling}$, however, is \emph{not} low-rank, so the literal low-rank Woodbury update of Eq.~(\ref{eq:woodbury}) provides no reduction and is quoted only as motivation. HPRMAT instead uses the equivalent factorization $\mathbf{M} = \mathbf{K}_{\rm diag}\,(\mathbf{I} + \mathbf{K}_{\rm diag}^{-1}\mathbf{V}_{\rm coupling})$, so that solving $\mathbf{M}\mathbf{c}=\mathbf{b}$ reduces to the Schur-type system $(\mathbf{I} + \mathbf{K}^{-1}\mathbf{V})\,\mathbf{c} = \mathbf{K}^{-1}\mathbf{b}$. The coupling therefore enters through the transformed operator $\mathbf{K}^{-1}\mathbf{V}$ rather than as a low-rank $\mathbf{U}\mathbf{C}\mathbf{V}$ correction; the Schur matrix and the transformed right-hand side are constructed explicitly in steps~3 and~4 below.}

The algorithm exploits this structure as follows:
\begin{enumerate}
    \item \textbf{Extract and invert kinetic blocks}: For each channel $\alpha$, extract the $n_{\rm lag} \times n_{\rm lag}$ kinetic energy block $\mathbf{K}_\alpha$ from the first channel (all channels share the same kinetic matrix). Compute $\mathbf{K}^{-1}$ using DGETRF/DGETRS (real arithmetic since kinetic energy is real);

    \item \textbf{Extract coupling potentials}: For each Lagrange point $r_i$, extract the $n_{\rm ch} \times n_{\rm ch}$ coupling matrix $\mathbf{V}(r_i)$ where $V_{\alpha\alpha'}(r_i) = M_{(\alpha-1)n_{\rm lag}+i, (\alpha'-1)n_{\rm lag}+i} - K_{ii}\delta_{\alpha\alpha'}$;

    \item \textbf{Build Schur complement}: Construct the $N \times N$ Schur complement matrix $\mathbf{S}$ with elements:
    \be
    S_{(i-1)n_{\rm ch}+\alpha, (j-1)n_{\rm ch}+\alpha'} = \delta_{ij}\delta_{\alpha\alpha'} + K^{-1}_{ij} V_{\alpha\alpha'}(r_j);
    \ee

    \item \textbf{Transform right-hand side}: Apply $\mathbf{K}^{-1}$ to the boundary vectors:
    \be
    \tilde{b}_{(i-1)n_{\rm ch}+\alpha} = \sum_{j=1}^{n_{\rm lag}} K^{-1}_{ij} q_2(r_j) \delta_{\alpha,\text{entrance}};
    \ee

    \item \textbf{Solve Schur system}: Solve $\mathbf{S}\tilde{\mathbf{x}} = \tilde{\mathbf{b}}$ using single-precision LU (CGETRF/CGETRS) for speed;

    \item \textbf{Extract R-matrix}: Transform solution back and compute R-matrix elements.
\end{enumerate}

Although the Schur complement has the same dimension $N \times N$ as the original system, this approach differs from the failed radial-diagonal Woodbury attempt (Section~\ref{sec:theory}) in two key ways: (1) the kinetic matrix $\mathbf{K}$ is real and shared across all channels, so $\mathbf{K}^{-1}$ is computed only once using efficient real arithmetic (DGETRF); (2) the Schur system is solved in single precision (CGETRF/CGETRS), which is approximately twice as fast as double precision on modern CPUs.

It is important to note that this solver does \emph{not} reduce the asymptotic complexity---the dominant cost remains $O(N^3)$ for the Schur complement solve. The performance gain is a \emph{constant-factor} improvement arising from several sources, which I quantify below.

\rev{\textbf{Cost comparison (Type~2 vs.\ Type~3).} A complex $N \times N$ LU factorization costs approximately $\frac{8}{3}N^3$ real floating-point operations \emph{independent of the working precision}: single precision (CGETRF) is faster than double precision (ZGETRF) not because it performs fewer operations but because each operation runs at roughly twice the throughput on a CPU (and up to $64\times$ on a consumer GPU). Type~2 performs this $\frac{8}{3}N^3$-operation factorization in single precision and adds an $O(N^2)$ iterative refinement. Type~3 involves: (i) one real $n_{\rm lag} \times n_{\rm lag}$ factorization, $\sim\frac{2}{3}n_{\rm lag}^3$ operations; (ii) Schur-complement construction, $O(N \cdot n_{\rm lag}^2)$ operations; (iii) a single-precision $N \times N$ factorization, the dominant $\frac{8}{3}N^3$ operations. The dominant cost is therefore the same single-precision factorization as Type~2; Type~3 can gain only a small constant-factor advantage because step (i) uses real arithmetic ($2\times$ faster than complex) and the kinetic inverse can be precomputed once and reused when the energy changes but the mesh is fixed. In practice the two are close: on the CPU-only platforms Type~3 is a few percent faster than Type~2 at large $N$ (Tables~\ref{tab:cpu_apple} and~\ref{tab:cpu_i9}), while on System~2 it is somewhat slower (Table~\ref{tab:gpu}); the trade-off is accuracy ($\sim 10^{-6}$ for Type~3 vs.\ $\sim 10^{-15}$ for Type~2). Type~3 is therefore recommended only when speed is critical and the problem does not involve ill-conditioned matrices (e.g., narrow resonances).}

\subsection{GPU cuSOLVER Solver (Type 4)}

The GPU solver uses NVIDIA's cuSOLVER library~\cite{cuSOLVER2024} to perform the linear solve on the GPU. Given the extreme FP32:FP64 performance ratio on consumer GPUs (64:1 on RTX 3090), I implement a mixed-precision strategy with all precision conversions performed on the GPU to minimize data transfer overhead.

The implementation uses persistent GPU memory allocation---buffers are allocated once and reused across multiple calls, avoiding the overhead of repeated allocation/deallocation. The algorithm proceeds as:
\begin{enumerate}
    \item \textbf{Initialize} (first call only): Create cuSOLVER and cuBLAS handles; query GPU properties; allocate persistent device memory for matrix ($N^2$ complex values), right-hand sides, pivot array, and workspace;

    \item \textbf{Host-to-device transfer}: Copy the double-precision matrix $\mathbf{M}$ and right-hand side $\mathbf{B}$ from CPU to GPU memory using cudaMemcpy;

    \item \textbf{FP64$\to$FP32 conversion on GPU}: Launch custom CUDA kernel \texttt{convert\_z2c\_kernel} to convert complex128 to complex64 entirely on the GPU:
    \begin{verbatim}
    __global__ void convert_z2c_kernel(
        const cuDoubleComplex* src,
        cuComplex* dst, int n) {
      int idx = blockIdx.x * blockDim.x + threadIdx.x;
      if (idx < n)
        dst[idx] = make_cuFloatComplex(
          (float)src[idx].x, (float)src[idx].y);
    }
    \end{verbatim}
    This avoids transferring data back to CPU for conversion;

    \item \textbf{Single-precision LU factorization}: Call \texttt{cusolverDnCgetrf} to compute the LU factorization $\mathbf{M}^{(s)} = \mathbf{P}\mathbf{L}^{(s)}\mathbf{U}^{(s)}$ on the GPU;

    \item \textbf{Single-precision solve}: Call \texttt{cusolverDnCgetrs} to solve the triangular systems;

    \item \textbf{FP32$\to$FP64 conversion on GPU}: Launch \texttt{convert\_c2z\_kernel} to convert the solution back to double precision;

    \item \textbf{Optional iterative refinement} \rev{(disabled by default): requesting refinement steps via the \texttt{max\_refine} parameter switches the solver to the host-refined hybrid mode described below. Each refinement step then:}
    \begin{enumerate}
        \item \rev{copies the current solution to the host and forms the residual $\mathbf{r} = \mathbf{b} - \mathbf{M}\mathbf{x}$ there in double precision (host BLAS ZGEMM), where FP64 runs at full rate;}
        \item \rev{checks convergence on the host (maximum residual modulus);}
        \item \rev{converts the residual to single precision and solves for the correction on the GPU using the stored LU factors;}
        \item \rev{invokes the custom kernel \texttt{add\_correction\_kernel} to add the correction to the device solution;}
    \end{enumerate}

    \item \textbf{Device-to-host transfer}: Copy the final solution back to CPU memory.
\end{enumerate}

The GPU memory layout uses column-major ordering (Fortran convention) to ensure compatibility with cuSOLVER. Error handling includes automatic fallback to CPU solver if GPU initialization fails or if cuSOLVER returns an error.

\rev{\textbf{Host-refined hybrid mode.} The refinement of step~7 is deliberately \emph{host}-refined. A fully on-device refinement would have to evaluate the residual $\mathbf{r} = \mathbf{b} - \mathbf{M}\mathbf{x}$ in double precision on the GPU, which on a consumer card runs at the throttled 1:64 rate; for the multi-channel systems of interest, where the number of right-hand sides grows with the channel count, this residual is costly enough to erode the factorization speedup. In the hybrid mode the single-precision factorization and triangular solves therefore run on the GPU, while the double-precision residual of each refinement step is formed on the host, where double precision runs at full rate. Only the right-hand-side-sized residual and correction vectors cross the PCIe bus at each step; the $N \times N$ matrix is transferred once. This recovers full double-precision accuracy on well-conditioned systems; on ill-conditioned ones the refinement still converges provided the single-precision factor is accurate enough to make each step a contraction, but it needs more steps as the condition number grows, while keeping the GPU factorization advantage. It is the recommended mode when double-precision accuracy is required on a consumer GPU.}

\rev{The crossover where the GPU becomes faster depends on the usage pattern. For a single isolated solve the one-time GPU initialization (context creation and device allocation) must be amortized, which keeps the CPU competitive below $N \approx 1000$. In steady-state operation, with the GPU context and buffers reused across an energy scan, the GPU is faster at all sizes tested and reaches about $15\times$ over the CPU direct solver (and $41\times$ over the legacy inversion) at $N = 25600$ on the RTX 3090.}

\textbf{GPU memory requirements.} \rev{The in-core GPU solver stores a double-precision copy of the matrix ($16N^2$ bytes), a single-precision working copy ($8N^2$ bytes), and the single-precision LU workspace. The latter is not negligible: the \texttt{cusolverDnCgetrf} buffer-size query returns one element per matrix entry (Lwork\,$=N^2$ verified at $N=25600$), i.e.\ a further $8N^2$ bytes; the solution and residual vectors add only $O(N\,n_{\rm ch})$. The footprint is therefore approximately $32N^2$ bytes. For a $24$~GB GPU this gives a recommended maximum of $N_{\max} \approx \sqrt{0.9 \times 24 \times 10^9 / 32} \approx 26000$, consistent with the largest case tested on the RTX~3090, $N = 25600$ (about $21$~GB). If the matrix exceeds available GPU memory, the solver automatically falls back to the CPU implementation. This $32N^2$ footprint applies to the in-core single-GPU path; the multi-GPU back-end of Section~\ref{sec:conclusion} keeps only the single-precision factorization on the device (its \texttt{cusolverMg} workspace is $O(N)$, not $O(N^2)$) and the double-precision matrix on the host, a leaner $\approx 8N^2$ per card that reaches larger $N$ even on one GPU.}

\section{Implementation}
\label{sec:implementation}

\subsection{Code Structure and API Compatibility}

HPRMAT is implemented in Fortran 90/95 with CUDA C extensions for GPU support. A primary design goal is to maintain API compatibility with Descouvemont's original R-matrix package~\cite{Descouvemont2016}. The main subroutine \texttt{rmatrix} preserves the same interface signature, argument order, and array conventions as the original code. This allows existing codes that call Descouvemont's package to use HPRMAT without any modification to their source code---only relinking is required.

The main components are:
\begin{itemize}
    \item \texttt{rmatrix\_hp.F90}: Main R-matrix interface with identical calling convention to Descouvemont's \texttt{rmatrix} subroutine;
    \item \texttt{rmat\_solvers.F90}: Four solver implementations with a unified internal interface;
    \item \texttt{gpu\_solver\_interface.F90}: Fortran-CUDA interface with runtime GPU detection;
    \item \texttt{cusolver\_interface.cu}: CUDA kernels for precision conversion and cuSOLVER calls;
    \item \texttt{special\_functions.f}: Coulomb functions (COULFG routine~\cite{Barnett1982}) and Whittaker functions (from FRESCO~\cite{Thompson1988});
    \item \texttt{angular\_momentum.f}: 3j, 6j, and 9j coefficients.
\end{itemize}

The only modification required in user codes is to select the solver type. This can be done in two ways:
\begin{enumerate}
    \item Setting the module variable \texttt{solver\_type} before calling \texttt{rmatrix}:
\begin{verbatim}
    use rmat_hp_mod
    solver_type = 4  ! 1=Dense, 2=Mixed, 3=Woodbury, 4=GPU
    call rmatrix(...)  ! Identical interface to Descouvemont's code
\end{verbatim}
    \item Passing the optional \texttt{isolver} argument directly to \texttt{rmatrix}:
\begin{verbatim}
    call rmatrix(..., isolver=4)  ! Select GPU solver for this call
\end{verbatim}
\end{enumerate}
If neither is specified, the default solver (Type 1, double-precision ZGESV) is used, providing identical results to Descouvemont's original code. This design ensures complete backward compatibility---existing codes work without any changes and automatically benefit from the optimized OpenBLAS implementation.

\subsection{BLAS Library Integration}

Performance of the CPU solvers depends critically on the underlying BLAS/LAPACK implementation. I recommend OpenBLAS~\cite{OpenBLAS}, which provides:
\begin{itemize}
    \item Multi-threaded BLAS operations optimized for modern CPUs;
    \item Support for all major architectures including x86\_64 and ARM (Apple Silicon);
    \item Consistent performance across operating systems.
\end{itemize}

\rev{The choice of OpenBLAS is a recommendation, not a requirement: HPRMAT calls only the standard BLAS/LAPACK interface, so it links against any compatible implementation. Users are free to substitute Intel MKL, AMD AOCL, or Apple Accelerate by relinking, with no change to the source code; OpenBLAS is used throughout this work because it is open-source and available on every platform tested (including Apple Silicon, where MKL is unavailable), which makes the benchmarks straightforward to reproduce.}

Thread management requires care to avoid nested parallelism conflicts. I recommend:
\begin{verbatim}
    export OPENBLAS_NUM_THREADS=<N>
    export OMP_NUM_THREADS=1
\end{verbatim}
where \texttt{<N>} is the number of physical cores \rev{(or available hardware threads)}.

\subsection{GPU Implementation}

The Fortran-CUDA interface uses ISO\_C\_BINDING for interoperability\rev{; for example, the single-GPU mixed-precision solver is bound as}:
\begin{verbatim}
    interface
      integer(c_int) function gpu_solve_mixed_c &
           (A, B, n, nrhs, max_refine, tol, info) &
           bind(C, name="gpu_solve_mixed_")
        use iso_c_binding
        complex(c_double_complex), intent(in)    :: A(*)
        complex(c_double_complex), intent(inout) :: B(*)
        integer(c_int), intent(in) :: n, nrhs, max_refine
        real(c_double), intent(in) :: tol
        integer(c_int), intent(out) :: info
      end function
    end interface
\end{verbatim}
\rev{with analogous bindings for the host-refined hybrid (\texttt{gpu\_solve\_hybrid\_}) and multi-GPU (\texttt{gpu\_solve\_multi\_mixed\_}) entry points. The matrix and right-hand sides are passed as double-precision complex arrays; all precision conversion takes place on the device (Section~\ref{sec:methods}).}

The CUDA implementation handles memory management, error checking, and cuSOLVER handle initialization. A fallback to CPU solvers is provided when no GPU is available or initialization fails.

\textbf{GPU memory requirements.} \rev{The device-memory footprint of the in-core GPU solver ($\approx 32N^2$ bytes) and the automatic fall-back when it is exceeded are quantified in the Type~4 description of Sec.~\ref{sec:methods}. As a further example, on an $8$~GB GPU such as the RTX~3070 the maximum practical size is $N \approx 15{,}000$.}

\subsection{Build System}

The build system consists of:
\begin{itemize}
    \item \texttt{setup.sh}: Auto-detection script for CUDA toolkit and OpenBLAS installation;
    \item \texttt{make.inc}: Configuration file generated by setup.sh or manually edited;
    \item \texttt{Makefile}: Modular build with optional GPU support.
\end{itemize}

Building with GPU support requires the NVIDIA CUDA Toolkit (version 11.5 or later). The GPU architecture flag (\texttt{sm\_XX}) should match the target GPU (e.g., \texttt{sm\_86} for RTX 3090).

\subsection{Language Bindings}

In addition to the native Fortran interface, HPRMAT provides bindings for C/C++, Python, and Julia, enabling integration with modern scientific computing workflows.

\subsubsection{C/C++ Interface}

The C interface uses Fortran's ISO\_C\_BINDING module to provide a clean C-compatible API. A header file \texttt{hprmat.h} declares all public functions:
\begin{verbatim}
    #include "hprmat.h"

    // Initialize
    double zrma[30];
    hprmat_init(30, 1, 10.0, zrma);

    // Set solver type
    hprmat_set_solver(HPRMAT_SOLVER_GPU);

    // Solve
    int nopen;
    double _Complex cu[nch * nch];
    hprmat_solve(nch, lval, qk, eta, rmax, nr, ns,
                 cpot, nr*ns, cu, &nopen, 0);
\end{verbatim}
The C interface is also usable from C++, Rust, Go, and other languages that support C foreign function interfaces.

\subsubsection{Python Interface}

The Python interface uses NumPy's f2py tool to generate a compiled extension module that calls Fortran directly without an intermediate C layer. A high-level \texttt{RMatrixSolver} class provides a Pythonic API:
\begin{verbatim}
    from hprmat import RMatrixSolver, SOLVER_GPU
    import numpy as np

    # Initialize solver
    solver = RMatrixSolver(nr=30, ns=1, rmax=10.0,
                           solver=SOLVER_GPU)

    # Set up problem
    cpot = np.zeros((30, nch, nch), dtype=np.complex128,
                    order='F')
    for ir, r in enumerate(solver.mesh):
        cpot[ir, 0, 0] = -50.0 * np.exp(-r**2 / 4.0)

    # Solve and get S-matrix
    S, nopen = solver.solve(lval, qk, eta, cpot)
\end{verbatim}
The Python wrapper handles array type conversion and memory layout (ensuring Fortran column-major order) automatically.

\subsubsection{Julia Interface}

The Julia interface uses \texttt{ccall} to invoke the Fortran library directly from a shared library (\texttt{libhprmat.so} or \texttt{libhprmat.dylib}). The \texttt{HPRMAT.jl} module provides a simple functional API that mirrors the Fortran subroutines:
\begin{verbatim}
    include("HPRMAT.jl")
    using .HPRMAT

    # Initialize - returns mesh points
    zrma = rmat_init(60, 1, 14.0)

    # Build potential (Julia is 1-indexed)
    cpot = zeros(ComplexF64, 60, nch, nch)
    for (i, r) in enumerate(zrma)
        cpot[i, 1, 1] = V(r) / hm
    end

    # Solve - returns S-matrix and nopen
    cu, nopen = rmatrix(nch, lval, qk, eta, rmax,
                        nr, ns, cpot, SOLVER_DENSE)
\end{verbatim}
Julia's native column-major array storage is directly compatible with Fortran, requiring no data layout conversion.

\section{Performance Results}
\label{sec:results}

I benchmark HPRMAT on three hardware platforms representing typical use cases: a high-end desktop CPU (Apple M3 Ultra), a workstation with consumer GPU (Intel Xeon + RTX 3090), and a mid-range CPU system (Intel i9-12900). All benchmarks use synthetic test matrices with the same block structure as physical R-matrix problems. \rev{This isolates solver performance and lets the matrix dimension be swept independently of any particular reaction; the physical validation of Fig.~\ref{fig:validation} confirms that the solvers reproduce the reference observables on real systems, so the synthetic timings are representative of production use.}

These three systems were chosen to represent the diversity of hardware available to potential users: System~1 demonstrates performance on modern ARM-based workstations (increasingly common in research environments), System~2 provides GPU benchmarks on a typical CUDA-capable workstation, and System~3 represents a conventional x86 desktop. While the absolute timings differ across systems, the \emph{relative} speedups of HPRMAT over the reference implementation are consistent, demonstrating that the algorithmic improvements are hardware-independent.

\subsection{Test Configuration}

\textbf{System 1 (CPU-only, Apple Silicon):}
\begin{itemize}
    \item CPU: Apple M3 Ultra (32 cores)
    \item Memory: 96 GB unified
    \item BLAS: OpenBLAS (multithreaded)\footnote{For this workload, OpenBLAS outperforms Apple's native Accelerate framework on Apple Silicon.}
\end{itemize}

\textbf{System 2 (CPU + GPU):}
\begin{itemize}
    \item CPU: \rev{$2\times$ }Intel Xeon Gold 6248R @ 3.00 GHz \rev{(48 physical cores total)}
    \item Memory: 630 GB
    \item GPU: NVIDIA GeForce RTX 3090 (24 GB, sm\_86)
    \item CUDA: 11.5
\end{itemize}

\textbf{System 3 (CPU-only, x86):}
\begin{itemize}
    \item CPU: Intel Core i9-12900 \rev{(16 cores, 24 threads)}
    \item Memory: 93 GB
    \item BLAS: OpenBLAS (OpenMP)\footnote{OpenBLAS was chosen over Intel MKL for reproducibility: it is open-source and available on all platforms (including Apple Silicon where MKL is unavailable). Modern OpenBLAS versions achieve performance within 5--10\% of MKL on Intel CPUs for dense complex linear algebra; the GPU speedups reported here would remain significant even with MKL as the CPU baseline.}
\end{itemize}

\rev{The GPU benchmarks were built with gfortran~9.4.0 and the CUDA~11.5 toolkit (device code compiled for compute capability \texttt{sm\_86}) and linked against OpenBLAS~0.3.8; the build scripts and the exact source revision accompany the public repository.}

The reference (``Ref.'' in tables) is Descouvemont's original R-matrix code~\cite{Descouvemont2016}, which uses \emph{explicit matrix inversion} via the LAPACK routine ZGETRI (compute $\mathbf{M}^{-1}$, then multiply $\mathbf{c} = \mathbf{M}^{-1}\mathbf{b}$). Crucially, to ensure that the observed speedups arise from algorithmic improvements and GPU acceleration rather than library differences, all CPU-based results (including the reference implementation) were obtained using the same optimized OpenBLAS library on each test system. \rev{All CPU timings, for both the reference and the HPRMAT solvers, were measured within a single process using identical BLAS threading: \texttt{OPENBLAS\_NUM\_THREADS} was set to 48 on System~2 (dual-socket, $2\times24$ physical cores), 32 on System~1 (Apple M3 Ultra, 32 cores), and 24 on System~3 (Intel i9-12900, all hardware threads), with \texttt{OMP\_NUM\_THREADS}\,$=$\,1, so the reference-versus-HPRMAT comparisons are strictly like-for-like. In the GPU benchmarks (Table~\ref{tab:gpu}), the CPU baselines (Types~1--3 and the reference) likewise used the fully multi-threaded OpenBLAS, so the reported GPU speedups are measured against fully-threaded, not single-core, CPU solvers.} The observed speedups therefore combine two effects: (1) the algorithmic advantage of direct solving (ZGESV) over explicit inversion (ZGETRI + ZGEMM), which contributes a factor of $\sim$1.5--2$\times$; and (2) mixed-precision and GPU acceleration, which provides the remaining speedup. Type~1 vs.\ Ref.\ isolates effect (1), while Type~2/3/4 vs.\ Ref.\ shows the combined improvement.

\subsection{CPU-Only Performance}

Table~\ref{tab:cpu_apple} shows benchmark results on System 1 (Apple M3 Ultra). All HPRMAT solvers significantly outperform the reference implementation, with speedups increasing for larger matrices.

\begin{table}[htbp]
\centering
\caption{Wall time (seconds) on Apple M3 Ultra (CPU only). Speedup relative to reference shown in parentheses.}
\label{tab:cpu_apple}
\begin{tabular}{l|c|c|c|c|c}
\hline
Matrix Size & Ref. & Type 1 & Type 2 & Type 3 & Best \\
\hline
$1024\times1024$ & 0.091 & 0.024 (3.7$\times$) & 0.036 (2.5$\times$) & 0.025 (3.6$\times$) & 3.7$\times$ \\
$2000\times2000$ & 0.489 & 0.077 (6.3$\times$) & 0.069 (7.1$\times$) & 0.060 (8.2$\times$) & 8.2$\times$ \\
$4000\times4000$ & 1.198 & 0.364 (3.3$\times$) & 0.338 (3.5$\times$) & 0.280 (4.3$\times$) & 4.3$\times$ \\
$8000\times8000$ & 8.162 & 2.066 (4.0$\times$) & 1.965 (4.2$\times$) & 1.319 (6.2$\times$) & 6.2$\times$ \\
$10000\times10000$ & 15.6 & 3.68 (4.2$\times$) & 3.05 (5.1$\times$) & 2.31 (6.8$\times$) & 6.8$\times$ \\
$16000\times16000$ & 60.5 & 14.1 (4.3$\times$) & 10.3 (5.9$\times$) & 8.82 (6.9$\times$) & 6.9$\times$ \\
$25600\times25600$ & 231.5 & 52.3 (4.4$\times$) & 34.0 (6.8$\times$) & 32.0 (7.2$\times$) & 7.2$\times$ \\
\hline
\end{tabular}
\end{table}

Type 3 (Woodbury-Kinetic) achieves the best performance on CPU for large matrices, reaching 7.2$\times$ speedup at $N = 25600$. Type 2 (Mixed Precision) provides comparable performance with higher accuracy.

Table~\ref{tab:cpu_i9} shows results on System 3 (Intel i9-12900), representing a typical high-end desktop configuration. Similar speedup trends are observed, with Type 2 and Type 3 achieving 5--6$\times$ speedup for large matrices.

\begin{table}[htbp]
\centering
\caption{Wall time (seconds) on Intel i9-12900 (CPU only, 24 threads).}
\label{tab:cpu_i9}
\begin{tabular}{l|c|c|c|c|c}
\hline
Matrix Size & Ref. & Type 1 & Type 2 & Type 3 & Best \\
\hline
$1024\times1024$ & 0.146 & 0.227 (0.6$\times$) & 0.020 (7.4$\times$) & 0.021 (6.9$\times$) & 7.4$\times$ \\
$2000\times2000$ & 0.294 & 0.393 (0.8$\times$) & 0.123 (2.4$\times$) & 0.114 (2.6$\times$) & 2.6$\times$ \\
$4000\times4000$ & 2.207 & 0.840 (2.6$\times$) & 0.539 (4.1$\times$) & 0.555 (4.0$\times$) & 4.1$\times$ \\
$8000\times8000$ & 21.9 & 7.04 (3.1$\times$) & 6.51 (3.4$\times$) & 4.76 (4.6$\times$) & 4.6$\times$ \\
$10000\times10000$ & 48.1 & 18.5 (2.6$\times$) & 12.1 (4.0$\times$) & 11.3 (4.3$\times$) & 4.3$\times$ \\
$16000\times16000$ & 163.0 & 54.7 (3.0$\times$) & 36.9 (4.4$\times$) & 32.9 (5.0$\times$) & 5.0$\times$ \\
$25600\times25600$ & 679.3 & 217.6 (3.1$\times$) & 125.1 (5.4$\times$) & 120.1 (5.7$\times$) & 5.7$\times$ \\
\hline
\end{tabular}
\end{table}

\subsection{GPU Performance}

Table~\ref{tab:gpu} shows benchmark results on System 2 (Intel Xeon + RTX 3090). The GPU solver (Type 4) provides dramatic speedups for large matrices.

\begin{table}[htbp]
\centering
\caption{\rev{Steady-state wall time (seconds) on Intel Xeon Gold 6248R + RTX 3090 (System 2). The GPU solver (Type 4) uses pinned host transfer, persistent buffers, and reads the host matrix directly (no defensive copy); all CPU solvers, including the reference, use OpenBLAS with \texttt{OPENBLAS\_NUM\_THREADS}\,$=$\,48 and \texttt{OMP\_NUM\_THREADS}\,$=$\,1, in a single shared process. Speedup relative to the reference (legacy explicit inversion) is shown in parentheses. The very large Type~4 speedups at $N=1000$--$2000$ are not a genuine GPU advantage: at those sizes the absolute times are dominated by the thread-dispatch overhead of multi-threaded BLAS on small matrices, which inflates the CPU (reference) timings. The production regime for the applications of Section~\ref{sec:intro} is $N \gtrsim 10^4$, where the Type~4 advantage is $\approx 15\times$ over the CPU direct solver (Type~1) and $\approx 41\times$ over the legacy inversion.}}
\label{tab:gpu}
\small
\begin{tabular}{l|c|c|c|c|c|c}
\hline
Matrix Size & Ref. & Type 1 & Type 2 & Type 3 & Type 4 (GPU) & Best \\
\hline
$1000\times1000$ & 1.061 & 1.109 (1.0$\times$) & 1.184 (0.9$\times$) & 1.201 (0.9$\times$) & 0.008 (134$\times$) & 134$\times$ \\
$2000\times2000$ & 1.508 & 1.516 (1.0$\times$) & 1.681 (0.9$\times$) & 1.770 (0.9$\times$) & 0.021 (70$\times$) & 70$\times$ \\
$4000\times4000$ & 2.765 & 2.405 (1.1$\times$) & 2.328 (1.2$\times$) & 2.623 (1.1$\times$) & 0.061 (46$\times$) & 46$\times$ \\
$8000\times8000$ & 9.306 & 5.597 (1.7$\times$) & 4.675 (2.0$\times$) & 5.087 (1.8$\times$) & 0.222 (42$\times$) & 42$\times$ \\
$10000\times10000$ & 14.409 & 7.347 (2.0$\times$) & 5.739 (2.5$\times$) & 6.515 (2.2$\times$) & 0.365 (40$\times$) & 40$\times$ \\
$16000\times16000$ & 40.103 & 15.700 (2.6$\times$) & 11.434 (3.5$\times$) & 15.748 (2.5$\times$) & 1.085 (37$\times$) & 37$\times$ \\
$25600\times25600$ & 144.948 & 52.098 (2.8$\times$) & 41.382 (3.5$\times$) & 48.067 (3.0$\times$) & 3.554 (41$\times$) & 41$\times$ \\
\hline
\end{tabular}
\end{table}

\rev{In steady-state operation, where the GPU context and device buffers are created once and reused across an energy scan, the Type~4 solver is faster than the CPU solvers at every size tested. At $N = 25600$ it completes a solve in $3.55$~s, versus $145$~s for the legacy inversion ($41\times$) and $52$~s for the optimized CPU direct solver ($15\times$). The advantage over the CPU direct solver is about $15$ to $20\times$ across the production range $N \gtrsim 10^4$ (and up to $25\times$ at $N = 8000$), largest at intermediate sizes and settling near $15\times$ at $N = 25600$, reflecting the ratio of single-precision GPU to double-precision multi-threaded-CPU throughput for this factorization; the larger ratio over the reference ($\approx 41\times$) is this factor multiplied by the algorithmic gain of direct solving over explicit inversion. For a single isolated solve the one-time GPU initialization (Section~\ref{sec:gpu_profile}) must instead be amortized, which makes a single CPU solve competitive below $N \approx 1000$. The very large speedups quoted for $N = 1000$--$2000$ in Table~\ref{tab:gpu} are an artifact of the multi-threaded CPU baseline, whose timings on such small matrices are dominated by BLAS thread-dispatch overhead, and should not be read as a genuine GPU advantage.}

\rev{It is instructive to compare the GPU and CPU on their shared single-precision factorization kernel, that is Type~4 against Type~2 with refinement switched off, so the two run the \emph{same} algorithm. At the production sizes the GPU is $11.6$ to $15.7\times$ faster than the multi-threaded CPU ($41.4$~s vs.\ $3.55$~s at $N = 25600$, i.e.\ $11.6\times$; $5.7$~s vs.\ $0.365$~s at $N = 10000$, i.e.\ $15.7\times$). On the CPU the two double-precision refinement steps of Type~2 are an $O(N^2)$ correction on top of the $O(N^3)$ factorization, so the tabulated Type~2 time is within a fraction of a percent of the bare single-precision factorization and serves here as the same-algorithm CPU reference. Why this same-algorithm ratio does not reach the nominal FP32 throughput ratio of the two devices is analysed by the profiling of Sec.~\ref{sec:gpu_profile}: once the host transfer is pinned, the cost is set mainly by the complex single-precision factorization. Type~2 adds two double-precision refinement steps on top of this factorization, which on the CPU are inexpensive and lift its accuracy to machine precision; the corresponding option on the GPU is the hybrid solver discussed next.}

\rev{\textbf{Double precision at GPU speed: the hybrid solver.}
The Type~4 timings above use single precision throughout, accurate to about $10^{-3}$ in the cross section, well within experimental uncertainties (Table~\ref{tab:accuracy}). When full double-precision accuracy is wanted, the host-refined hybrid mode delivers it on the same consumer card. Table~\ref{tab:hybrid} compares the two on the well-conditioned synthetic matrices: two refinement steps recover machine-precision agreement with the double-precision reference (maximum solution error $\sim 10^{-16}$, versus $\sim 10^{-10}$ for single precision) at a cost of about $2.5$ to $3\times$ the single-precision time. The number of refinement steps needed, and hence the cost, grows with the conditioning of the matrix. On the physical $\alpha+{}^{208}$Pb validation (Fig.~\ref{fig:validation}(c)) the deviation from the reference falls by about a factor of five per step, so two steps reach $\sim 10^{-5}$ and machine precision takes several more; in every case the result is orders of magnitude below experimental precision. At the two-step setting the hybrid remains $5.3\times$ faster than the optimized double-precision CPU solver (Type~1) at $N = 25600$ ($9.85$~s vs.\ $52$~s), and $5$ to $7.5\times$ faster across $N \gtrsim 10^4$. The two modes are complementary: single precision for maximum throughput where it suffices, the hybrid where higher accuracy is wanted.}

\begin{table}[htbp]
\centering
\caption{\rev{Single-precision (Type~4) versus host-refined hybrid GPU solver on System~2 (RTX~3090), steady state, for the well-conditioned synthetic benchmark matrices (hybrid with two refinement steps). Time in seconds with speedup over the legacy inversion reference in parentheses; ``Max error'' is the maximum relative solution-vector error against the double-precision reference. On ill-conditioned physical systems more refinement steps are needed to reach the same accuracy (Fig.~\ref{fig:validation}(c)).}}
\label{tab:hybrid}
\rev{
\begin{tabular}{r|cc|cc}
\hline
 & \multicolumn{2}{c|}{Type~4 (single precision)} & \multicolumn{2}{c}{Hybrid (double precision)} \\
$N$ & Time (speedup) & Max error & Time (speedup) & Max error \\
\hline
$10000$ & $0.365$ ($40\times$) & $7\times10^{-10}$ & $0.98$ ($15\times$) & $2\times10^{-16}$ \\
$16000$ & $1.085$ ($37\times$) & $3\times10^{-10}$ & $3.10$ ($13\times$) & $6\times10^{-17}$ \\
$25600$ & $3.554$ ($41\times$) & $8\times10^{-10}$ & $9.85$ ($15\times$) & $6\times10^{-16}$ \\
\hline
\end{tabular}
}
\end{table}

\textbf{Note on timing methodology.} Wall times in Table~\ref{tab:gpu} include complete GPU operations: host-to-device matrix transfer, LU factorization, triangular solves, iterative refinement (if applicable), and device-to-host result transfer. This reflects realistic usage where the matrix is updated on the CPU at each energy point and transferred to the GPU for solving. GPU memory for workspace arrays is allocated once at initialization and reused across multiple calls, which is the recommended usage pattern for energy scans.

\begin{figure}[htbp]
\centering
\includegraphics[width=0.9\textwidth]{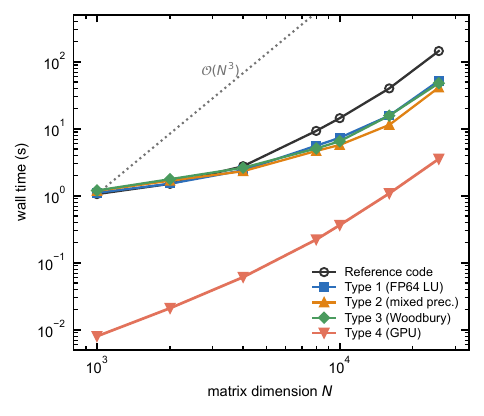}
\caption{Scaling of computation time with matrix dimension $N$ (log-log scale). The dotted line is an $O(N^3)$ slope guide anchored at the smallest matrix size; the solvers approach this asymptotic slope at large $N$, while the shallower behavior at small $N$ reflects fixed per-call and thread-dispatch overhead. Note: Absolute timings are hardware-dependent; this plot shows results on a specific test system (Intel Xeon Gold 6248R + NVIDIA RTX 3090) and is intended to illustrate the relative performance of different solvers and the scaling behavior, not absolute performance on other hardware.}
\label{fig:scaling}
\end{figure}

Figure~\ref{fig:scaling} shows the scaling of computation time with matrix size on a log-log plot. The GPU solver (Type 4) demonstrates the expected $O(N^3)$ scaling while maintaining a consistent advantage over CPU solvers for large matrices. The increasing speedup with matrix size reflects the amortization of GPU overhead costs.

\subsection{\rev{GPU Kernel Profiling}}
\label{sec:gpu_profile}

\rev{To understand what limits the same-algorithm GPU-to-CPU ratio (Type~4 vs.\ Type~2), I instrumented the steady-state Type~4 solve at $N = 25600$ on the RTX~3090 with per-stage device timers; Table~\ref{tab:profile} gives the breakdown of the optimized solver. The profiling first exposed two host-side costs that the GPU kernels themselves do not incur. (i) The solver wrapper made a defensive copy of the $N\times N$ matrix into a work buffer before each solve; at $N = 25600$ this $10.5$~GB copy ($\approx 21$~GB of host memory traffic) took $\approx 1.9$~s. (ii) The host-to-device transfer of the double-precision matrix moves the $16N^2 \approx 10.5$~GB operand across PCIe; with pageable host buffers it ran at only $\approx 5$~GB/s, taking $\approx 2.1$~s. The single-precision factorization (\texttt{cusolverDnCgetrf}) itself takes $\approx 2.5$~s and is the single dominant stage; the FP32 conversion kernels, the triangular solves, and the device-to-host transfer of the solution together contribute under $5\%$ (Table~\ref{tab:profile}). The factorization was therefore not the original bottleneck: the two host-side data movements were.}

\rev{Guided by this profile I made two changes that remove both host-side costs and leave the solve factorization-bound. First, I page-lock (pin) the host matrix so its transfer uses direct DMA instead of a staged pageable copy; the registration is cached and the host work buffers are persistent, so in an energy scan the one-time page-locking is paid only at the first solve. This assumes the standard usage in which a single host matrix buffer is reused across the scan: that buffer must remain allocated between successive GPU solves and be released through the library's finalize entry before it is freed, a condition the energy-scan workflow satisfies automatically. This raises the effective bandwidth to $\approx 12$~GB/s and cuts the per-solve transfer from $\approx 2.1$~s to $\approx 0.86$~s. Second, I eliminate the defensive matrix copy on the GPU path entirely: the device holds its own working copy and the GPU solver never writes back to the host matrix (a writable copy is needed only by the CPU fallback), so the solver now reads the caller's matrix directly. With both changes the steady-state solve at $N = 25600$ falls from $5.5$~s to $3.55$~s, of which the FP32 factorization is $2.5$~s (Table~\ref{tab:profile}): the solve is now factorization-bound, as intended, and the GPU timings in Table~\ref{tab:gpu} reflect both optimizations. The transferred volume could be halved once more by converting the matrix to single precision on the host before transfer, a natural next optimization since iterative refinement is disabled by default (Section~\ref{sec:methods}); this applies to the pure single-precision Type~4 path only, since the host-refined hybrid mode keeps the double-precision matrix on the host to form the residual.}

\begin{table}[htbp]
\centering
\caption{\rev{Per-stage breakdown of the steady-state Type~4 (GPU single-precision) solve at $N = 25600$ on the RTX~3090, after pinning the host transfer and removing the redundant host-side matrix copy. The one-time GPU context creation and host page-locking ($\approx 1.1$~s) are paid once per energy scan and are not included.}}
\label{tab:profile}
\rev{
\begin{tabular}{l|c}
\hline
Stage & Time (s) \\
\hline
Host-to-device transfer (pinned, $10.5$~GB) & $0.86$ \\
FP64\,$\to$\,FP32 conversion & $0.02$ \\
FP32 LU factorization (\texttt{cusolverDnCgetrf}) & $2.49$ \\
FP32 triangular solves (\texttt{cusolverDnCgetrs}) & $0.08$ \\
FP32\,$\to$\,FP64 conversion + device-to-host & $0.02$ \\
Host RHS assembly + R-matrix extraction & $0.07$ \\
\hline
Total (steady state) & $3.54$ \\
\hline
\end{tabular}
}
\end{table}

\subsection{Accuracy Validation}

Table~\ref{tab:accuracy} summarizes the accuracy of each solver. The ``Max Error'' column reports the maximum relative error in the solution vector $\|\mathbf{x}_{\rm test} - \mathbf{x}_{\rm ref}\|_\infty / \|\mathbf{x}_{\rm ref}\|_\infty$ compared to Type~1 as reference. The ``Physics Error'' column reports the corresponding error in computed cross sections, which is the quantity relevant for practical applications.

\begin{table}[htbp]
\centering
\caption{Accuracy of different solvers on well-conditioned systems. ``Max Error'' is the maximum relative error in the solution vector; ``Physics Error'' is the typical relative error in computed cross sections. \rev{For the iterative-refinement solvers (Type~2 and the hybrid) these figures assume convergence; the number of refinement steps needed grows with the matrix conditioning, as shown for a physical case in Fig.~\ref{fig:validation}(c). For Type~4 the two columns refer to different test problems: the $\sim 10^{-10}$ solution-vector error is measured on the well-conditioned synthetic matrices, whereas the $\sim 10^{-3}$ cross-section figure is the conservative value on physical, less well-conditioned systems, so the ``physics error below solution error'' relation noted in the text holds only within a single system. The Max Error entries are representative precision floors at mid-range $N$; they are essentially $N$-independent except for Type~3, whose error decreases with $N$ (from $\sim 5\times10^{-5}$ at $N=1000$ to $\sim 7\times10^{-8}$ at $N=25600$; see text).}}
\label{tab:accuracy}
\small
\begin{tabular}{l|c|c|l}
\hline
Solver & Max Error & Physics Error & Description \\
\hline
Type 1 (OpenBLAS ZGESV) & $\sim 10^{-18}$ & --- & Machine precision (reference) \\
Type 2 (Mixed Precision) & \rev{$\sim 10^{-15}$} & $<10^{-14}$ & Double prec.\ (refined) \\
Type 3 (Woodbury-Kinetic) & $\sim 10^{-6}$ & $<1\%$ & Nuclear physics accuracy \\
Type 4 (GPU cuSOLVER) & $\sim 10^{-10}$ & \rev{$\sim 10^{-3}$} & \rev{GPU FP32, no refinement by default} \\
\rev{Hybrid (GPU + host refine)} & \rev{$\sim 10^{-16}$} & \rev{$<10^{-14}$} & \rev{GPU FP32 factor.\ + host FP64 refinement} \\
\hline
\end{tabular}
\end{table}

The physics error is generally smaller than the solution vector error because cross sections depend on the solution through a weighted sum (the R-matrix element), which averages out random numerical errors. For Type~3, the $\sim 10^{-6}$ solution error translates to sub-percent cross-section errors, which is well within typical experimental uncertainties in nuclear physics ($\gtrsim 1\%$).

\rev{The accuracy is essentially independent of the matrix dimension. Across the benchmark range $N = 1000$ to $25600$ the maximum error stays at the precision floor of each algorithm, with no growth in $N$: $\sim 10^{-18}$ for Type~1, $\sim 10^{-15}$ for Type~2, and $\sim 10^{-10}$ for Type~4. Type~3 in fact improves with $N$, from $\sim 5\times 10^{-5}$ at $N=1000$ to $\sim 7\times 10^{-8}$ at $N=25600$, as the single-precision Schur solve benefits from the larger, well-conditioned kinetic block. These tests use well-conditioned matrices; the conditioning dependence is treated in the paragraph below.}

\textbf{Conditioning and resonance behavior.} R-matrix calculations near narrow resonances or at deep sub-barrier energies can produce ill-conditioned matrices. \rev{To assess robustness, I tested the solvers on matrices with condition numbers up to $\kappa \sim 10^{8}$ (typical of narrow-resonance regions). Type~2 (CPU mixed precision with iterative refinement) maintained accuracy to $\sim 10^{-10}$ for $\kappa < 10^{6}$; for $\kappa \sim 10^{8}$ it degraded to $\sim 10^{-6}$, still sufficient for cross-section calculations. The GPU solver (Type~4) runs the FP32 factorization and solve without iterative refinement by default, which is accurate to $\sim 10^{-10}$ for the well-conditioned systems benchmarked here; for ill-conditioned problems its refinement should be enabled (the \texttt{max\_refine} option), restoring the Type~2 behavior at the cost of additional FP64 residual evaluations. Type~3 (single precision without refinement) showed larger errors ($\sim 10^{-3}$) for $\kappa > 10^{6}$ and is not recommended near very narrow resonances. In such cases Type~1 (full double precision), or Type~2 / Type~4 with refinement enabled, should be used.}

\begin{figure}[htbp]
\centering
\includegraphics[width=0.95\textwidth]{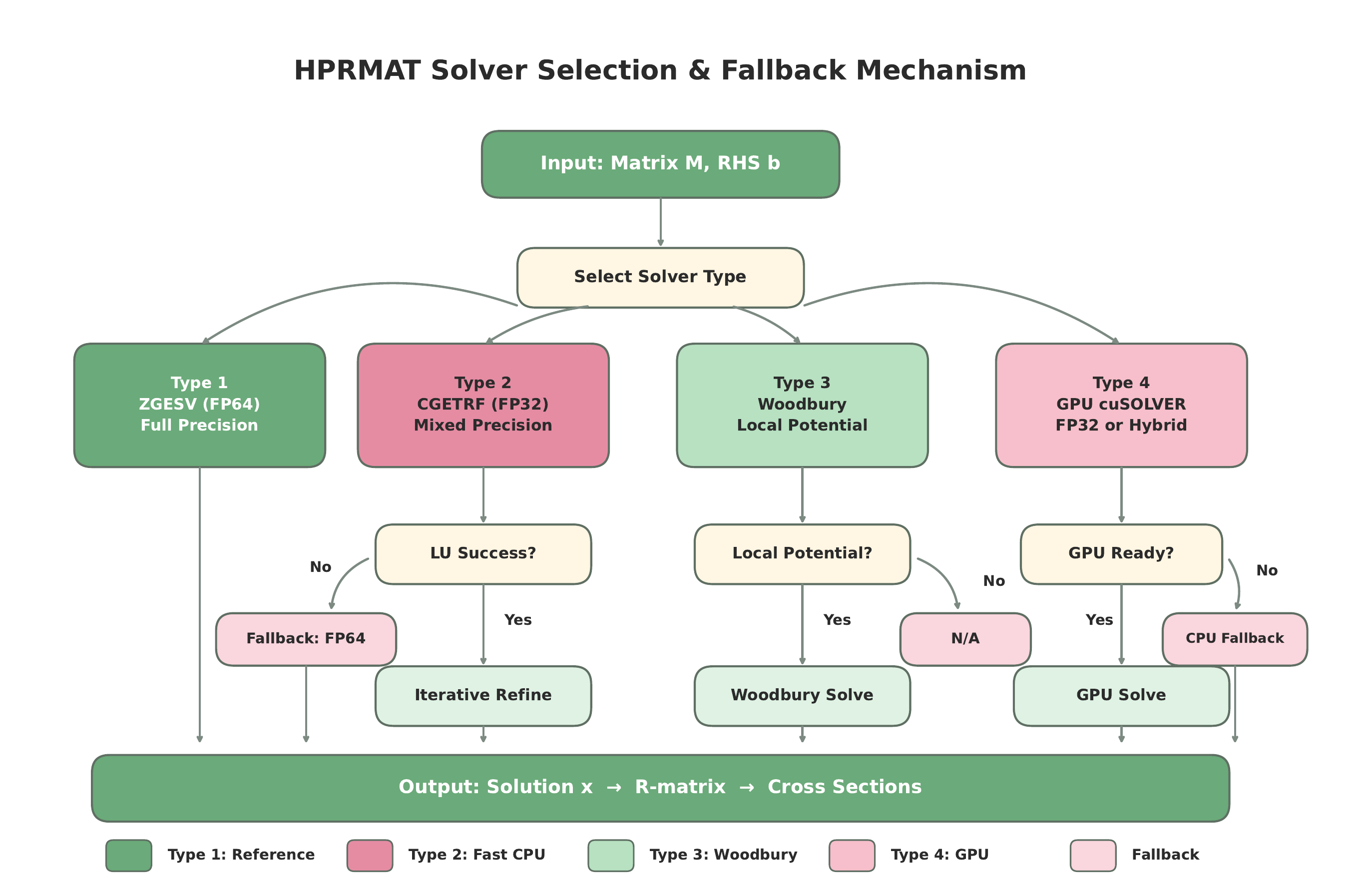}
\caption{HPRMAT solver selection and fallback mechanism. Each solver type has automatic fallback paths to ensure numerical safety: Type~2 falls back to FP64 if single-precision LU fails; Type~4 falls back to CPU solvers if GPU is unavailable or cuSOLVER fails. Type~3 (Woodbury) requires local potentials and is not applicable for non-local interactions. \rev{The GPU solver (Type~4) runs in single precision by default; when full double-precision accuracy is required it switches to the host-refined hybrid mode, in which the double-precision residual of each refinement step is formed on the CPU (Section~\ref{sec:methods}).}}
\label{fig:flowchart}
\end{figure}

\textbf{Fallback mechanism.} HPRMAT includes automatic fallback to ensure numerical safety. If the single-precision LU factorization fails (e.g., due to singular or near-singular matrices), the solver automatically falls back to full double-precision ZGESV. Similarly, if GPU initialization or cuSOLVER execution fails, the code transparently falls back to CPU solvers. Users working with extremely ill-conditioned problems ($\kappa > 10^{8}$, as may occur near very narrow resonances with widths $\Gamma \lesssim 1$~eV) should use Type~1 (full FP64) to guarantee maximum numerical stability. Figure~\ref{fig:flowchart} illustrates the solver selection and fallback logic.

\subsection{Physical Validation}

\begin{figure}[htbp]
\centering
\includegraphics[width=0.9\linewidth]{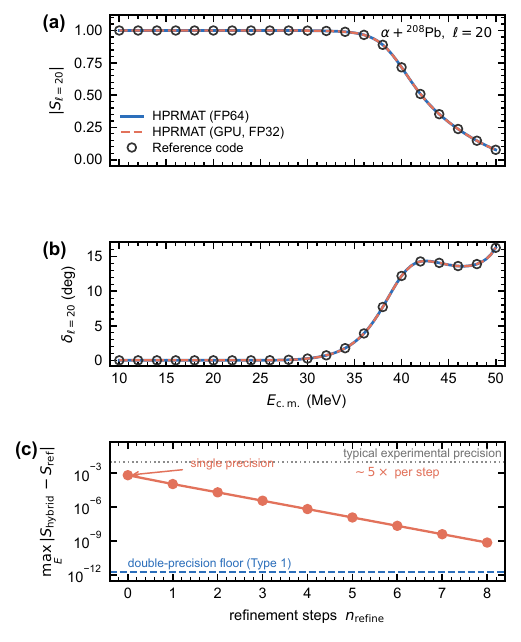}
\caption{\rev{Validation of HPRMAT against Descouvemont's reference R-matrix code on a physical observable: the $\ell = 20$ collision-matrix element for $\alpha+{}^{208}$Pb elastic scattering (Goldring optical potential). (a) modulus $|S|$ and (b) phase shift $\delta$ as functions of the c.m.\ energy; the reference (open circles), the HPRMAT double-precision solver (solid), and the HPRMAT GPU single-precision solver (dashed) are visually indistinguishable. (c) Convergence of the GPU hybrid solver: the maximum deviation from the reference over the energy range falls geometrically (about a factor of five per step) with the number of host refinement steps, from the single-precision value at zero steps ($6.8\times10^{-4}$) toward the double-precision floor set by the reference comparison ($\sim 10^{-12}$, dashed line). The deviation is below the $\sim 10^{-2}$ scale of typical experimental uncertainties (dotted line) at every refinement level, including single precision.}}
\label{fig:validation}
\end{figure}

I validate HPRMAT against Descouvemont's reference code using all five standard test cases from his R-matrix package~\cite{Descouvemont2016}:

\textbf{$\alpha$+$^{208}$Pb optical model} (1 channel): Elastic scattering with the Goldring potential. \rev{Figure~\ref{fig:validation} compares the $\ell = 20$ collision-matrix element $S(E)$ computed with HPRMAT and with the reference code over $E_{\rm c.m.} = 10$--$50$ MeV, spanning the grazing region where $|S|$ falls from unity to $\approx 0.08$. The double-precision solver (Type~1) reproduces the reference to machine precision (maximum deviation $1.8\times 10^{-12}$); the GPU single-precision solver (Type~4) agrees to $6.8\times 10^{-4}$, more than an order of magnitude below the $\sim 1\%$ scale of typical experimental uncertainties. Figure~\ref{fig:validation}(c) shows the convergence of the GPU hybrid mode on this system: each host refinement step reduces the deviation by about a factor of five, from $6.8\times 10^{-4}$ at zero steps toward the double-precision floor, reaching $\sim 10^{-9}$ by eight steps. The number of steps needed is set by the conditioning of the matrix, so this physical system converges more slowly than the well-conditioned synthetic matrices used for Table~\ref{tab:hybrid}; even so, two steps already reach $\sim 10^{-5}$, orders of magnitude below experimental precision.}

\textbf{Nucleon-nucleon scattering} (2 channels): Reid soft-core potential with $T=1$. Phase shifts agree to better than $10^{-4}$ degrees.

\textbf{$^{16}$O+$^{44}$Ca coupled-channel} (4 channels): Inelastic scattering with rotational coupling. Cross sections agree within 0.1\%.

\textbf{$^{12}$C+$\alpha$ inelastic scattering} (12 channels): Scattering amplitudes agree within 0.1\% for Types 1, 2, and 4, and within 1\% for Type 3. This level of agreement is well within typical experimental uncertainties.

\textbf{Non-local Yamaguchi potential} (1 channel): Tests the non-local potential capability. Phase shifts agree to machine precision for Type 1.

\subsection{Solver Selection Guidelines}

Based on my benchmarks, I recommend:
\begin{itemize}
    \item \textbf{GPU available}: Use Type 4 for $N > 1000$;
    \item \rev{\textbf{GPU available, double-precision accuracy required}: Use Type 4 in its host-refined hybrid mode (enabled by requesting a nonzero number of refinement steps), which recovers double-precision accuracy (machine precision after a few refinement steps on well-conditioned systems, and more steps on ill-conditioned ones) while remaining several times faster than the CPU double-precision solver;}
    \item \textbf{CPU only, large matrices, local potentials}: Use Type 3 (Woodbury) for best speed, or Type 2 (Mixed Precision) for higher accuracy;
    \item \textbf{Non-local potentials}: Use Type 1, 2, or 4 (Type 3 is not applicable);
    \item \textbf{High precision required}: Use Type 1 (ZGESV)\rev{, or Type 4 in hybrid mode for GPU speed};
    \item \textbf{Validation/debugging}: Use Type 1 for reference.
\end{itemize}

\section{Summary and Outlook}
\label{sec:conclusion}

I have presented HPRMAT, \rev{a self-contained, high-performance R-matrix solver framework for coupled-channel scattering calculations. It provides the full propagation machinery with the same user-supplied-potential interface as standard R-matrix packages, and is additionally drop-in compatible with the linear algebra routines of Descouvemont's package.} To my knowledge, HPRMAT represents the current \emph{state-of-the-art} in R-matrix solver performance for nuclear physics, being the first publicly available implementation to combine GPU acceleration, mixed-precision arithmetic, and structure-exploiting algorithms within a unified framework. \rev{The solver backends themselves depend only on the block structure of Eq.~(\ref{eq:block_structure})---dense kinetic-plus-Bloch diagonal blocks with diagonal local-coupling off-diagonal blocks---which is common to every microscopic Lagrange-mesh R-matrix code in nuclear, atomic, and molecular physics; the drop-in compatibility with Descouvemont's package is a deployment convenience layered on top of solvers that are in principle transferable to any code producing this matrix class.}

A key motivation for this work is \emph{democratizing high-performance computing} in nuclear physics. Data-center GPUs (NVIDIA A100, H100) offer excellent FP64 performance but cost \$10,000--\$30,000 and are often inaccessible to university research groups. Consumer GPUs (RTX 3090/4090/5090) cost \$1,000--\$2,000 and are readily available, but their FP64 performance is deliberately limited (64:1 ratio vs.\ FP32). HPRMAT's mixed-precision strategy transforms this limitation into an advantage: the bulk of the computation runs in single precision, which for nuclear reaction observables already meets the required accuracy (about $10^{-3}$ in the cross section, well below experimental uncertainties), \rev{and when full double precision is wanted the host-refined hybrid solver recovers it on the same consumer card, still several times faster than a multi-threaded CPU.} Researchers can thus run large-scale CDCC calculations on desktop workstations that previously required cluster access. The key innovations are:
\begin{itemize}
    \item Adoption of direct linear solvers (standard in numerical computing but historically overlooked in legacy nuclear codes), providing numerical stability improvements and modest performance gains;
    \item GPU acceleration achieving \rev{about 15$\times$ speedup over the optimized CPU direct solver (41$\times$ over the legacy inversion-based code) on a consumer-grade RTX 3090, an advantage that would be even higher on more powerful consumer GPUs (e.g., RTX 4090, RTX 5090), which offer superior FP32 throughput compared to data-center GPUs (I have not benchmarked those cards, but the single-precision-bound factorization is expected to scale with their FP32 throughput)}, \rev{with an optional host-refined hybrid mode that recovers double-precision accuracy in a few refinement steps, several times faster than the optimized double-precision CPU solver;}
    \item Mixed-precision algorithms exploiting the favorable FP32:FP64 performance ratio on modern hardware;
    \item \rev{A tested multi-GPU back-end that distributes the single-precision factorization across several consumer cards, extending the reachable matrix dimension well beyond a single card (to $N=76{,}800$ on four RTX~3090s) in full double precision recovered by host-side iterative refinement, with per-card device memory scaling as $\approx 1/P$ at large $N$;}
    \item Woodbury formula optimization exploiting the kinetic-coupling block structure of the R-matrix Hamiltonian (for local potentials);
    \item Multi-language support with C, Python, and Julia bindings for seamless integration into modern scientific workflows.
\end{itemize}

All solvers have been validated against Descouvemont's reference implementation~\cite{Descouvemont2016} using all five standard test cases from his package. The code is designed to be a drop-in replacement for existing R-matrix calculations, requiring minimal changes to user codes.

\rev{\textbf{Scalability and multi-GPU support.} A natural question is whether problems beyond the single-card memory limit require distribution across several GPUs. The limit depends on where the double-precision matrix is held. The single-GPU solver benchmarked above keeps the double-precision matrix, a single-precision copy, and the single-precision LU workspace on the device (the $\approx 32\,N^2$-byte model of Section~\ref{sec:methods}), giving $N_{\max}\approx 26{,}000$ on a $24$~GB card ($N=25{,}600$ tested). The multi-GPU back-end uses a leaner split: it stores only the single-precision factorization on the device ($8$~bytes per element, with an $O(N)$ \texttt{cusolverMg} workspace) and keeps the double-precision matrix on the host, where it is needed solely to form the iterative-refinement residual. With this footprint even a single card reaches $N=51{,}200$ in full double precision (peak $20.9$~GB; residual $\sim 3\times10^{-14}$ against a manufactured exact solution; the one-GPU column of Table~\ref{tab:mgpu}). This already exceeds the physically motivated regime: as quantified in Section~\ref{sec:intro}, even a demanding XCDCC calculation for an extended halo such as $^{11}$Be reaches $N\sim 10{,}000$ to $30{,}000$, so the realistic range fits one consumer card with headroom.}

\rev{For problems larger still, the back-end distributes this single-precision factorization across the visible GPUs with \texttt{cusolverMg} in a one-dimensional block-cyclic column layout, the double-precision residual again formed on the host (the multi-GPU analogue of the single-GPU hybrid solver); it is selected at run time like the other back-ends, as solver type~6 of the \texttt{rmatrix}/\texttt{isolver} interface of Section~\ref{sec:implementation}. Because the per-card storage is the distributed factor, the per-card device memory scales as roughly $1/P$ at large $N$; the reduction is imperfect because a per-card overhead (the \texttt{cusolverMg} workspace and CUDA context, of order $1$~GB on these runs) does not distribute, so the smaller-$N$ rows of Table~\ref{tab:mgpu} shrink much less than $1/P$. Table~\ref{tab:mgpu} reports wall time, accuracy, and per-card peak memory on a $4\times$ RTX~3090 system (the same card as System~2): the per-card requirement for $N=51{,}200$ falls from $20.9$~GB on one card to $11.2$~GB on two and $6.3$~GB on four, which lets the solver reach $N=64{,}000$ on two cards and $N=76{,}800$ on four, all in full double precision (residual $\sim 6\times10^{-14}$ on these well-conditioned test matrices). Because the back-end keeps the double-precision matrix resident on the host to form the residual, that $16N^2$ host copy ($\approx 94$~GB at $N=76{,}800$) is the one store distribution cannot shrink: per-card device memory falls as cards are added (the dashes in Table~\ref{tab:mgpu} mark where it exceeds the $24$~GB card), whereas the host matrix is irreducible and, for the $630$~GB of System~2, would bind only near $N\approx 2\times10^{5}$, far beyond the device-limited sizes tabulated here. The wall-time gain from added cards is modest (for $N=51{,}200$: $53.1$, $44.7$, and $42.9$~s on one, two, and four cards) because the double-precision residual that recovers full accuracy is formed on the host and is not distributed; a pure double-precision distribution parallelizes better but is far slower overall, since it runs the whole factorization at the throttled consumer-card FP64 rate. The multi-GPU benefit is therefore capacity and per-card memory, not raw speed. This is consistent with the central message of the work: a single consumer card covers the physics, while the multi-GPU path is the documented, benchmarked route for the occasional problem that exceeds one card, without recourse to data-center resources.}

\begin{figure}[htbp]
\centering
\includegraphics[width=0.85\textwidth]{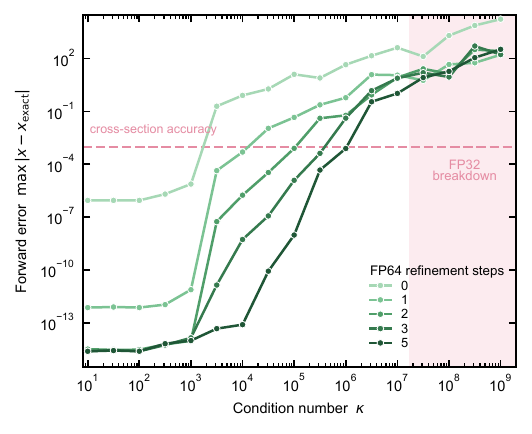}
\caption{\rev{Accuracy of the mixed-precision multi-GPU solver as a function of the matrix condition number $\kappa$, for a Hermitian test matrix of prescribed $\kappa$ (built by a Householder similarity of a graded diagonal) with manufactured exact solution $\mathbf{x}_{\rm exact}=\mathbf{1}$ ($N=8000$, two RTX~3090 GPUs). Each curve is the forward error $\max|\mathbf{x}-\mathbf{x}_{\rm exact}|$ for a fixed number ($0$ to $5$) of double-precision host-refinement steps, with the production convergence fallback disabled so that the intrinsic accuracy of the refined single-precision solve is exposed. Each refinement step lowers the attainable error and extends the double-precision plateau to higher $\kappa$. Beyond $\kappa \sim 10^{7}$ (shaded; $\kappa\,u_{\rm FP32}\sim 1$, with $u_{\rm FP32}$ the single-precision unit roundoff) the single-precision factor is too inaccurate for the refinement to converge and all curves rise to $O(1)$; in deployment the solver detects this from the residual and falls back to a full-precision solve rather than returning such a result. The dashed line marks the $\sim 10^{-3}$ accuracy sufficient for cross sections.}}
\label{fig:illcond}
\end{figure}

\rev{The ``full double precision'' of the mixed-precision path is conditioning-dependent, exactly as for any single-precision factorization with iterative refinement. Figure~\ref{fig:illcond} maps the forward error against the matrix condition number $\kappa$ for a manufactured exact solution: each host-refinement step lowers the attainable error and pushes the double-precision plateau to higher $\kappa$, so the default two steps recover machine precision for well-conditioned systems, while additional steps extend the machine-precision plateau to $\kappa \sim 10^{4}$ and keep the error below the cross-section requirement up to $\kappa \sim 10^{6}$. Beyond $\kappa \sim 10^{7}$ (where the product of $\kappa$ and the single-precision unit roundoff $u_{\rm FP32}\approx 6\times10^{-8}$ approaches unity) the single-precision factor is too inaccurate for the refinement to converge. With refinement enabled (the default for this path) the solver detects the non-convergence from the residual and falls back to a full-precision solve, so it does not return a single-precision-limited result silently; the bare single-precision mode, selected by requesting zero refinement steps, is an explicit speed-over-accuracy choice and carries no such guard. For the narrow-resonance and deep sub-barrier regimes that can produce such ill-conditioning, the double-precision Type~1 solver (or Type~2 with refinement) remains the appropriate choice, as discussed in Section~\ref{sec:results}.}

\begin{table}[htbp]
\centering
\caption{\rev{Mixed-precision multi-GPU solver (single-precision \texttt{cusolverMg} factorization, double-precision host refinement with two refinement steps) on a $4\times$ NVIDIA RTX~3090 ($24$~GB) system with dual Xeon Gold~6248R (the System~2 card). For each matrix dimension $N$ and number of GPUs, the steady-state wall time $t$, the peak per-card device memory, and the forward error against a manufactured exact solution ($\mathbf{x}_{\rm exact}=\mathbf{1}$, so $\mathbf{b}=A\mathbf{1}$) are shown; the right-hand side has $64$ columns. A dash indicates a configuration whose per-card memory exceeds the $24$~GB limit. The per-card memory scales as $\approx 1/P$ with the number of GPUs $P$, so distribution lifts the reachable $N$ from $51{,}200$ on one card to $76{,}800$ on four, in full double precision. Benchmarked with CUDA~11.5; the distributed factorization uses the \texttt{cusolverMg} multi-GPU interface.}}
\label{tab:mgpu}
\begin{tabular}{r|cc|cc|cc|c}
\hline
 & \multicolumn{2}{c|}{1 GPU} & \multicolumn{2}{c|}{2 GPUs} & \multicolumn{2}{c|}{4 GPUs} & \\
$N$ & $t$\,(s) & mem\,(GB) & $t$\,(s) & mem\,(GB) & $t$\,(s) & mem\,(GB) & $\max|\mathbf{x}-\mathbf{x}_{\rm exact}|$ \\
\hline
$12{,}800$ & $2.2$  & $2.0$  & $2.2$  & $1.4$  & $2.4$   & $1.1$  & $1.9\times10^{-14}$ \\
$25{,}600$ & $10.7$ & $5.9$  & $9.7$  & $3.5$  & $9.8$   & $2.2$  & $2.5\times10^{-14}$ \\
$38{,}400$ & $25.7$ & $12.2$ & $22.5$ & $6.7$  & $23.4$  & $4.0$  & $4.4\times10^{-14}$ \\
$51{,}200$ & $53.1$ & $20.9$ & $44.7$ & $11.2$ & $42.9$  & $6.3$  & $3.5\times10^{-14}$ \\
$64{,}000$ & ---    & ---    & $81.2$ & $16.9$ & $76.2$  & $9.3$  & $6.9\times10^{-14}$ \\
$76{,}800$ & ---    & ---    & ---    & ---    & $112.2$ & $12.8$ & $6.1\times10^{-14}$ \\
\hline
\end{tabular}
\end{table}

\section*{CRediT authorship contribution statement}
\textbf{Jin Lei:} Writing -- review \& editing, Writing -- original draft, Visualization, Validation, Software, Resources, Project administration, Methodology, Investigation, Funding acquisition, Formal analysis, Conceptualization.

\section*{Declaration of generative AI and AI-assisted technologies in the manuscript preparation process}
During the preparation of this work the author used Claude (Anthropic) in order to polish the manuscript text and debug program code. After using this tool, the author reviewed and edited the content as needed and takes full responsibility for the content of the published article.

\section*{Declaration of competing interest}
The author declares no known competing financial interests or personal relationships that could have appeared to influence the work reported in this paper.

\section*{Acknowledgements}
This work was supported by the National Natural Science Foundation of China (Grant Nos.~12475132 and 12535009) and the Fundamental Research Funds for the Central Universities.

\section*{Data availability}
No data was used for the research described in the article. The program files are available in the CPC Library at \url{https://doi.org/10.17632/gcnpfnz3w6.1} and in the developer's repository at \url{https://github.com/jinleiphys/HPRMAT}.

\bibliographystyle{elsarticle-num}
\bibliography{inclusive_prc}
\end{document}